# OPMOS: Ordered Parallel Algorithm for Multi-Objective Shortest-Paths

Leo Gold
University of Connecticut
Storrs, CT USA
leo.gold@uconn.edu

Adam Bienkowski
University of Connecticut
Storrs, CT USA
adam.bienkowski@uconn.edu

David Sidoti
US Naval Research Laboratory
Monterey, CA USA
david.m.sidoti.civ@us.navy.mil

Krishna Pattipati
University of Connecticut
Storrs, CT USA
krishna.pattipati@uconn.edu

Omer Khan
University of Connecticut
Storrs, CT USA
khan@uconn.edu

## Abstract

The Multi-Objective Shortest-Path (MOS) problem finds a set of Pareto-optimal solutions from a start node to a destination node in a multi-attribute graph. The literature explores multi-objective A*-style algorithmic approaches to solving the NP-hard MOS problem. These approaches use consistent heuristics to compute an exact set of solutions for the goal node. A generalized MOS algorithm maintains a "frontier" of partial paths at each node and performs ordered processing to ensure that Pareto-optimal paths are generated to reach the goal node. The algorithm becomes computationally intractable at a higher number of objectives due to a rapid increase in the search space for non-dominated paths and the significant increase in Pareto-optimal solutions. While prior works have focused on algorithmic methods to reduce the complexity, we tackle this challenge by exploiting parallelism to accelerate the MOS problem. The key insight is that MOS algorithms rely on the ordered execution of partial paths to maintain high work efficiency. The proposed parallel algorithm (OPMOS) unlocks ordered parallelism and efficiently exploits the concurrent execution of multiple paths in MOS. Experimental evaluation using the NVIDIA GH200 Superchip's 72-core Arm-based CPU shows the performance scaling potential of OPMOS on work efficiency and parallelism using a real-world application to ship routing.

## 1 Introduction

In many optimization problems, several distinct (and often competing) objectives need to be considered simultaneously. For example, when planning a road trip, one may wish to minimize the driving distance, driving time, and cost of tolls along the route. Similarly, when planning a journey by sea, one may be interested in the fastest and the most fuel-efficient routes, but deciding the right trade-off may depend on the urgency of the matter and the meteorological and oceanographic (METOC) environment.

This paper explores the NP-hard multi-objective shortest-path (MOS) problem, a generalization of the well-known (and polynomial) single-objective shortest-path problem [27]. Given a weighted graph with non-negative edge weights, the shortest path problem computes the minimum-cost path from a start node to a goal/destination node in the graph [32]. In a multi-objective setting, each edge is given a non-negative cost vector (constant length for each edge in a graph), with each element corresponding to an objective. When these objectives compete, no single path can optimize all the objectives simultaneously. MOS aims to find a set of Pareto-optimal (non-dominated) solution paths, where a path is Pareto-optimal if no single objective of the path can be improved without causing at least one of the other objectives to deteriorate in quality. For example, (5,4) and (4,5) can be Pareto-optimal path costs to a node. During execution, the intermediate path cost vectors form the so-called *Pareto-optimal labels*[1] that comprise potential candidate solutions. However, computing this front is computationally hard [3, 5, 20, 27], even for two objectives [7]. As the number of objectives increases, so does the computational complexity and the number of Pareto-optimal solution paths [24].

---

[1] MOS literature also identifies a candidate path as a label [16, 24, 26].

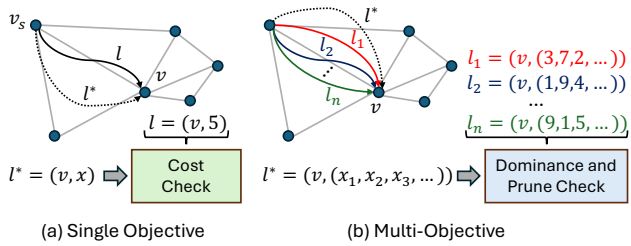

**Figure 1: Comparison of (a) single objective vs. (b) multi-objective intermediate labels between the source node $v_s$ and an intermediate node $v$. A new candidate label $l^*$ is shown alongside the associated label-level operations.**

To efficiently compute the Pareto-optimal solution front, algorithmic solutions are developed based on the multiobjective extension of the A* algorithm designed for single-objective search [15, 31, 34]. Multi-objective A* (MOA*), unlike A* that exits once the first solution is found, needs to store a set of Pareto-optimal solution paths to the goal node. A New Approach to Multi-Objective A* (NAMOA*) [15] uses consistent heuristics for the A* search to ensure that an exact set of solution paths are computed to the goal node. NAMOA* handles an arbitrary number of objectives to compute an exact set of solution paths, thus serving as the basis for the proposed MOS algorithm in this paper.

In a single-objective shortest path, there can be only one minimum solution path cost for each node in the graph. As seen in Figure 1(a), this is computed using a singular cost comparison for a given node in the graph. However, MOS does not have a single solution path guarantee since multiple non-dominated paths can exist from the start node to any other node in the graph. If a label $l(v)$ is defined to be a path cost from the source node to a node $v$, when a new candidate label $l^*(v)$ is discovered, a *dominance check* must be performed to verify if the accumulated cost vector for any previous labels for that node $v$ dominate the cost vector for $l^*(v)$. The label $l^*(v)$ must be compared with all previously found non-dominated labels to $v$, as illustrated in Figure 1(b). If it is not dominated, then the mutually non-dominated set of labels at $v$ must be updated with the new label, possibly *pruning* existing solutions if the new label dominates them. These dominance and pruning checks are expensive, especially as the number of non-dominated labels increases with the number of objectives. A recent survey paper [25] suggests key challenges facing MOS algorithms. These include tackling the computational complexity of label processing with increasing objectives, choosing the best lexicographic order when ordering the candidate labels, and creating scalable and efficient parallel MOS implementations.

Identifying the right labels for parallelization is crucial to unlocking parallel performance potential in MOS. In the sequential setting, a lexicographically ordered priority ensures a globally optimal label is extracted at each iteration [15]. However, when multiple labels are extracted in a parallel setting, the resulting labels may not be Pareto-optimal. Sanders and Mandow [26] reach the same conclusion in their theoretical parallel model for one of the original bi-criteria MOS algorithms [16]. Their approach focuses on extracting the global Pareto front at each iteration using a specialized Pareto-queue data structure confined to only two objectives. They assert that an exact implementation of the global Pareto queue is unknown for more than two objectives and, consequently, propose a controlled relaxation of label ordering for more than two objectives. This forms the basis for our proposed hypothesis: the work efficiency of MOS needs to be kept in check by extracting labels as close to global Pareto-optimal ordering as possible. With increasing objectives, the priority queue presents more labels that may be close to being globally Pareto-optimal. If high-priority labels are extracted, there exists the potential for a lower reduction in work efficiency since each wasted label processed is a complex operation with a high order of dominance and pruning checks. The key idea of *ordered queue extraction* is proposed for parallel MOS. It encompasses two main components: candidate label ordering and the number of labels extracted at each iteration.

We make the following observations from a performance characterization of MOS: (1) the complexity of the algorithm grows substantially with the number of objectives, (2) there is an opportunity for exploiting parallelism given the large time spent in candidate label computations relative to the priority queue operations, (3) load balanced work distribution is challenging due to extreme variability in work performed for each candidate label, and (4) the ordering of the candidate labels is salient for work efficiency. Herein, an Ordered Parallel MOS algorithm (OPMOS) is proposed to capture the importance of maintaining order in computations while extracting parallel performance for MOS acceleration. It aims to keep the order intact for extracting high-priority labels from a lexicographically-ordered priority queue (PQ) of intermediate non-dominated labels. These labels are distributed among the cores of a multicore CPU processor at each iteration. OPMOS faces load balancing and the inherent serialization of PQ operations as key performance scaling challenges.

The work done for each label is not uniform since intermediate graph connections are explored in an unstructured manner, and each label performs a non-deterministic number of dominance and pruning checks (unique to MOS). The challenge worsens as the graph size and the number of objectives increase. A label-complexity-aware load-balancing method is proposed to reduce the load imbalance in OPMOS. When multiple labels are processed in parallel, the label extractions and updates to the PQ must be performed sequentially. To overcome the serialization bottleneck, an asynchronous execution of parallel label processing is proposed to hide the latency of PQ operations. This paper makes the following contributions:

- The NAMOA* MOS algorithm is evaluated using the Tool for Multi-objective Planning and Asset Routing (TMPLAR) [30] that deploys a set of state-space reduction techniques to generate graphs with over 10 objectives. The performance characterization reveals several core observations: the growth in complexity with an increase in objective count, the potential for parallelism due to long computational critical paths, the extreme variability in the work done for each candidate label, and the importance of ordering candidate labels for work-efficient parallel execution.

- OPMOS is proposed for parallel MOS execution that handles an arbitrary number of objectives to compute an exact set of solution paths. OPMOS is evaluated to highlight the work efficiency and parallelism trade-offs. The evaluation using a 72-core Arm CPU shows a geometric mean 34× speedup over sequential for the evaluated graphs using the NAMOA* algorithm.

## 2 Related Work

The MOS problem is well-studied from an algorithmic perspective, and it is known that generating an exact Pareto-front is NP-hard [27]. Alternative genetic and evolutionary algorithms [2, 12, 38, 42] have been explored in the literature, but they suffer from computational inefficiencies and poor explainability of the quality of their solutions. Therefore, researchers have focused on tackling the algorithmic complexity of the generative Pareto front approaches, as summarized in a recent survey paper [25]. Algorithmic techniques have been explored to reduce the runtime complexity or approximate the Pareto front using label-setting or label-correcting methods. Martin's algorithm [16] is a label-setting algorithm that extends single-objective Dijkstra to the multi-objective setting. MOA* [31] introduces A* to the multi-objective domain. Since then, many improvements over MOA* have been explored in the literature, with NAMOA* [15] serving as the basis of most modern advancements [25]. Algorithmic enhancements include: dimensionality reduction (NAMOA*-dr [23]), lazy versus eager dominance checks (BOA* [34]), and enhanced data structures (EMOA* [24]), among others [25].

While these algorithmic optimizations have been proposed, most focus on two or three objectives due to the exploding size of the search space induced by the increasing number of objectives [24, 25]. NAMOA* remains one of the only modern MOS algorithms that applies to an arbitrary number of objectives, establishing itself as the baseline for this paper. Due to the NP-hard nature of the exact generative algorithms, approximations to the Pareto front have also been explored to lower the complexity at higher objective counts through runtime state-space reductions. Warburton [37] introduces an $\epsilon$-based procedure that allows approximate dominance checks to enable pruning of paths within an $\epsilon$-bounded range. Several optimizations to the approximation strategy have been introduced [3, 4, 9, 33]. The quality of solutions is impacted with approximations, introducing a trade-off between runtime efficiency and solution quality [40]. In this paper, we retain the exact solution quality of NAMOA* and propose a parallel NAMOA* algorithm to tackle its computational complexity.

So far, all discussed strategies address the MOS complexity from an algorithmic perspective. Parallel MOS is an under-explored method of handling the complexity as the number of objectives increases [25]. Sanders and Mandow [26] present a parallel variant of Martin's algorithm in the bi-objective setting. It constructs a Pareto queue to allow the parallel extraction of all globally Pareto-optimal labels at each iteration, which the authors assert is not practical for more than two objectives. Focusing on theoretical analysis, this paper does not introduce an implementation or experiments and notes that the proposed algorithm may not be practical. Others attempt parallelization by launching multiple MOS instances with different lexicographical orderings of objectives [25]. However, there is no known single-instance parallel MOS algorithm in the literature that is capable of handling an arbitrary number of objectives. We propose to address the MOS complexity problem for higher numbers of objectives using ordered parallelism.

A key challenge in systematically evaluating MOS algorithms is the lack of multi-objective graph benchmarks using real-world applications [25]. Prior works use synthetic graphs to constrain the search space or use a road network with only two or three objectives [24]. The search space for MOS in the New York City map (264$K$ nodes and 733$K$ edges) is so massive that limited evaluation is performed in [24] by comparing partial solution paths obtained within a runtime limit of 600 seconds. TMPLAR [17, 30, 40] is the only framework (to the best of our knowledge) that evaluates real-world maritime ship routing. It uses a spatio-temporal setting and a set of state-space reduction techniques to generate graphs with over 10 objectives using a variety of dynamic weather and ship datasets, presenting real-world scenarios requiring multi-objective path planning. TMPLAR uses the NAMOA* MOS algorithm to produce ship routes based on an arbitrary number of input objectives. Despite the state-space reduction, many routes are still intractable for high numbers of objectives. Therefore, the maximum number of objectives completed within a predetermined runtime limit is used for evaluation in this paper.

Ordered graph processing is the cornerstone of extracting parallelism in modern graph applications [8, 21]. Various concurrent priority schedulers for graph analytics have been introduced in the literature, ranging from hardware-centric to software approaches. Swarm [13] and its variant Hive [22] propose speculative execution of tasks in hardware to achieve super-linear speedups for task-parallel graph problems. HD-CPS [28] proposes a hardware-software co-design to trade off work efficiency and parallelism in concurrent priority schedulers for task parallel graph processing. Many CPU and GPU frameworks have also been proposed for ordered graph processing, such as MBQ [39], Galois [18], GraphIt [41], and Gunrock [36]. Dynamic load balancing schedulers are also explored to overcome challenges related to work irregularity in graph problems [1, 14]. While all these ordered graph frameworks and enhancements show promise, no work has pursued these methods for the MOS problem. Our proposed approach to exploiting ordered parallelism in MOS can fit into any ordered graph processing framework. However, we propose a general-purpose ordered parallelization of the MOS problem in this paper.

## 3 Background and Complexity of MOS

Consider an input graph $G = (V, E, c)$ with a set of nodes $V$ and edges $E$. For each edge $e \in E$, there is a non-negative cost vector $c(e)$ of length $d$ objectives. Given a source node $v_s$ and a goal node $v_g$ in the graph $G$, the path from $v_s$ to an intermediate node $v_i$ is defined as $\pi(v_i)$, represented by a sequence of nodes where each node is connected to its predecessor on the path. For each path $\pi(v_i)$, $\hat{g}(\pi(v_i))$ denotes the path cost from $v_s$ to $v_i$, calculated as the sum of the cost vectors $c(e)$ for all edges present on the path. Since multiple objectives may compete, MOS introduces a dominance check such that given two paths $a = \pi_1(u), b = \pi_2(u)$ with $d$ objectives, $a$ dominates $b$ (denoted $a \succeq b$) if and only if $\hat{g}(a)[i] \leq \hat{g}(b)[i], \forall i \in 1, 2, ..., d$, and $\hat{g}(a)[i] < \hat{g}(b)[i], \exists i \in 1, 2, ..., d$. All non-dominated paths from $v_s$ to $v_g$ constitute the *Pareto-optimal* solution set. MOS aims to find a *cost-unique* Pareto-optimal solution set where no two paths in the subset have the same cost vector.

A few additional terms must be introduced to describe the MOS problem. A label $l = (v, \hat{g})$ is a tuple representing an intermediate solution path from $v_s$ to $v \in V$ with a cost vector $\hat{g}$. For simplicity, we denote $v(l)$ to be the vertex and $\hat{g}(l)$ to be the cost vector contained in $l$. A label $l$ dominates another label $l'$ ($l \succeq l'$) if they share the same vertex ($v(l) = v(l')$) and $\hat{g}(l) \succeq \hat{g}(l')$. A heuristic vector $\hat{h}(v)$ is an *admissible* heuristic such that it dominates (less than or equal for all objectives) all Pareto-optimal solutions from node $v$ to the goal node [15]. A vector $\hat{F}(l)$ denotes the estimated path cost from the start node to the goal node for a given label, calculated as $\hat{F}(l) = \hat{g}(l) + \hat{h}(v(l))$. Let OPEN be a queue of labels prioritized by $\hat{F}(l)$ in increasing lexicographic order. For each vertex $u \in V$, let $\alpha(u)$ denote the *frontier* set at node $u$, holding all non-dominated labels $l$ at node $u$. Each label in $\alpha(u)$ is a non-dominated partial solution path from $v_s$ to $u$. In NAMOA*, $\alpha$ is split into two sets $G_{OP}$ and $G_{CL}$, the open and closed sets, respectively. Here, $G_{OP}$ contains a per-node set of all partial solution labels in OPEN, while $G_{CL}$ contains the remaining non-dominated solution labels in the frontier set of each node. Every label in $\alpha(u)$ can be found in either $G_{OP}(u)$ or $G_{CL}(u)$ for all nodes $u \in V$. NAMOA* also maintains the *Pareto-optimal solution front*, $P$, holding the frontier set at the goal node ($\alpha(v_g)$). The output of NAMOA* is an *exact* set of Pareto-optimal solution paths in $P$.

As shown in Algorithm 1, after initializing the data structures (lines 1-2), a label for the start node $l_s = (v_s, \hat{0})$ is created and inserted into both OPEN (with priority $\hat{0}$) and $G_{OP}(v_s)$ (lines 3-4). At each iteration (lines 6-31), the label

---

**Algorithm 1** Sequential NAMOA*

**Input:** Edge costs $C$, heuristics $H$, start $v_s$ and goal $v_g$ nodes
1: OPEN ← PriorityQueue($\emptyset$); $P \leftarrow \emptyset$
2: $G_{op}(u) \leftarrow \emptyset, G_{CL}(u) \leftarrow \emptyset, \forall u \in V$
3: $l_s \leftarrow (v_s, \hat{0})$
4: OPEN.insert($l_s$), $G_{OP}(v_s)$.insert($l_s$)
5: **while** OPEN $\neq \emptyset$ **do**
6:    $l \leftarrow$ OPEN.popmin()
7:    $G_{OP}(v(l))$.remove($l$); $G_{CL}(v(l))$.insert($l$)
8:    **if** $v(l) = v_g$ **then**
9:       *PrunedOPENLabels* ← PruneOPEN($l$)
10:      **for all** $l^* \in$ *PrunedOPENLabels* **do** $G_{OP}(v(l^*))$.remove($l^*$)
11:      Prune($P, l$)
12:      **if** NotDominated($l, P$) **then**
13:         $P$.insert($l$)
14:    **else**
15:      **for all** $v' \in$ GetNeighbors($v(l)$) **do**
16:         $l' \leftarrow (v', \hat{g}(l) + \hat{c}(v(l), v');$ parent($l' \leftarrow l$)
17:         $\hat{F}(l') \leftarrow \hat{g}(l') + \hat{h}(v(l'))$
18:         **if not** Visited($v'$) **then**
19:            **if** NotDominated($\hat{F}(l'), P$) **then**
20:               OPEN.insert($l'$)
21:               $G_{OP}(v(l'))$.insert($l'$)
22:         **else if** Duplicate($l'$) **then**
23:            continue
24:         **else if** NotDominated($l', G_{OP}(v(l'))$) **and**
25:              NotDominated($l', G_{CL}(v(l'))$) **then**
26:            Prune($G_{CL}(v(l')), l'$)
27:            *PrunedLabels* ← Prune($G_{OP}(v(l')), l'$)
28:            **for all** $l^* \in$ *PrunedLabels* **do** OPEN.remove($l^*$)
29:            **if** NotDominated($\hat{F}(l'), P$) **then**
30:               OPEN.insert($l'$)
31:               $G_{OP}(v(l'))$.insert($l'$)
32: **return** $P$

with the lexicographically lowest $\hat{F}$-vector is extracted from OPEN (line 6). This label $l$ is removed from $G_{OP}$ and inserted into $G_{CL}$ (line 7) to update the frontier sets. At this point, there are a few key procedures to define.

**NotDominated (l, S)** compares $l$ with labels in a given set $S$ to verify if a label exists in $S$ that dominates $l$. It returns *false* if $l$ is dominated by a label in $S$, and returns *true* otherwise.

**Prune (S, l)** searches through all labels in a given set $S$, and removes all labels in $S$ that are dominated by $l$.

**PruneOPEN (l)** searches the entire OPEN and removes all labels in OPEN heuristic-dominated by $l$ (i.e. $\hat{F}(l) \geq \hat{F}(l^*)$ for an $l^* \in$ OPEN).

If $l$ is at the goal node ($v(l) = v_g$, line 8), PruneOPEN is called (line 9) to prune out labels heuristic-dominated by $l$ in OPEN. These labels are subsequently removed from $G_{OP}$ (line 10). This operation requires a computationally expensive full index search through OPEN every time the goal node is reached. The complexity of this operation grows substantially with the number of objectives due to a rapid increase in the number of candidate labels. Then, Prune is called (line 11) to remove labels in $P$ dominated by $l$, ensuring the $P$ set contains only globally Pareto-optimal solution labels. NotDominated is called to check if $l$ is not dominated by any labels in $P$ (line 12), and if successful $l$ is inserted into the Pareto-optimal solution front $P$ (line 13). The complexity of this operation is proportional to the number of candidate labels in $P$, which grows with the number of objectives.

If $l$ is *not* at the goal node, then all neighbors $v'$ of $v(l)$ are explored (line 15). For each $v'$, a new candidate label $l' = (v', \hat{g}(l) + \hat{c}(v(l), v'))$ is generated by extending $l$ from $v(l)$ to $v'$. The parent pointer of $l'$ is set to $l$ to allow solution path reconstruction once execution concludes (line 16). The new $\hat{F}(l')$ is also computed, combining the new path cost $\hat{g}(l')$ with the heuristic cost $\hat{h}(v(l'))$ (line 17) to create a lower-bound estimate of the total path cost from the source $v_s$ to the goal node $v_g$. If $v'$ is explored for the first time (line 18), then extra computations can be skipped under the assumption that $G_{OP}(v')$ and $G_{CL}(v')$ are empty. Before $l'$ can be inserted into OPEN and $G_{OP}$ (lines 20-21), NotDominated is called (line 19) to check if the lower-bound path estimate ($\hat{F}(l')$) is dominated by any goal-node solutions in $P$. The complexity of this operator is also proportional to candidate labels in $P$.

If $v'$ has been visited, then $l'$ is compared against the labels in $G_{OP}(v(l'))$ and $G_{CL}(v(l'))$ in procedure Duplicate($l'$) to check if $l'$ is a duplicate label, and if it is, the rest of the iteration is skipped (lines 22-23). Otherwise, dominance checks are performed to see if any label in $G_{OP}(v(l'))$ or $G_{CL}(v(l'))$ dominates $l'$ (lines 24-25). The complexity of the duplicate and dominance checks depends on the number of candidate labels in $G_{OP}$ and $G_{CL}$ for this node, which grows with the number of objectives. If $l'$ is not dominated, all labels from the frontier set of $v(l')$ dominated by $l'$ are pruned (lines 26-27). The labels pruned from $G_{OP}$ are memorized to prune them from OPEN and avoid a full index search (line 28). Before $l'$ can be inserted into OPEN and $G_{OP}$ (lines 30-31), NotDominated is called (line 29) to check if the lower-bound path estimate ($\hat{F}(l')$) is dominated by any goal-node solutions in $P$. Again, the complexity of the dominance check operator is proportional to candidate labels in the set searched ($P$, $G_{OP}$ and $G_{CL}$ for this node). The Prune operators' complexity also increases with candidate labels being processed for a given node, which grows with the number of objectives.

Once OPEN is empty, the algorithm terminates, and $P$ contains the final labels representing the Pareto-optimal solution paths (also referred to as exact solutions). Each label processed during the MOS execution performs unstructured and nondeterministic work. Depending on the graph characteristics, each label processed may explore an arbitrary number of adjacent labels. For each neighbor label, a nondeterministic number of labels must be compared via dominance and pruning checks. The complexity of these operators relies on the number of candidate labels being compared, which increases with the number of objectives. Given the high complexity of each label processed, NAMOA* emphasizes reducing the work being performed. The OPEN is implemented as a Priority Queue with lexicographical ordering of objectives, guaranteeing that a Pareto-optimal label is extracted at each iteration. This ensures that a candidate label with the highest chance of remaining in the final solution is processed, reducing redundant work.

## 4 Characterization and Motivation

Many representative graph datasets, such as road networks, become infeasible for MOS to handle due to the rapid explosion of the state space as the number of objectives increases [24]. TMPLAR attempts to solve this challenge using a variety of graph state-space reduction techniques [30, 40]. A forward and backward single-source shortest path (SSSP) is performed to compute a bounding box of reachable nodes to reduce the search space while also creating a time expansion of the graph to account for weather conditions over time. Then, the edge weights, followed by an *admissible* heuristic (using SSSP), are computed for each objective. TMPLAR generates directed spatio-temporal graphs with >10 objectives. The details about the objectives (Table 1) and routes (Table 2) are discussed in Section 6. However, to characterize the computational challenges, the MOS Algorithm 1 is evaluated with Route 1 (cf. Table 2), as used in [40].

Figure 2 (left) shows the measured execution time and the number of labels extracted from OPEN with increasing objective counts from two to the maximum objectives. The OPEN extractions are established as a metric of the work performed and algorithmic complexity. As the number of objectives increases, the execution time increases rapidly, as does the relative work performed. The execution time of 3 milliseconds is recorded for two objectives, which rises to over 38 minutes for 12 objectives. For more insights, Figure 2 (right) shows the runtime breakdown split into label computations and OPEN (PQ) operations for two to twelve objectives. At lower objectives, less time is spent on label processing due to the low complexity of label operations. With few labels per node at lower objectives, few dominance checks are performed per candidate label. As the number of objectives increases, more candidate labels exist, resulting in an explosion in label processing complexity. At higher objectives, the runtime is dominated by label processing, and the priority queue operations become less prominent.

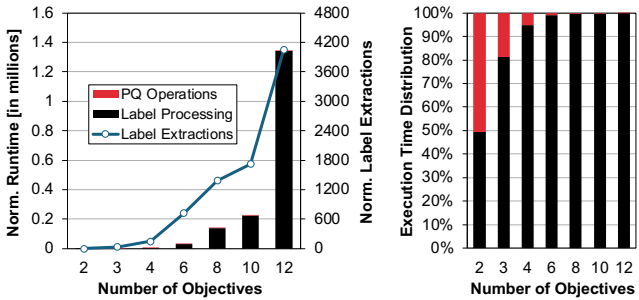

Figure 2: TMPLAR Route 1 sequential MOS (left) relative runtime performance and label extractions normalized to two objectives, and (right) execution time distribution for two to twelve objectives.

The updates for each label can be performed before extracting a new label from OPEN, decoupling the extraction/updates from the label computations. If multiple labels can be processed simultaneously, these independent computations can be parallelized to improve runtime efficiency. However, exploiting label-level parallelism is not trivial. Figure 3 shows the number of label checks and comparisons per label processed for two, six, and twelve objectives. If some nodes have many candidate labels, they require more checks than nodes with few labels. Each label performs tens to hundreds of comparisons at two objectives, which increases to hundreds of thousands to millions of comparisons at twelve objectives. This highlights the high variability in label computations. The irregular nature of label processing presents significant load-balancing challenges for MOS. Determining how to distribute these labels for parallel execution becomes more important as the number of objectives increases.

Another key factor for processing labels in parallel is extracting multiple labels in a given iteration. Diverging from the global Pareto-optimal order may result in redundant labels, leading to work inefficiency. A single global Pareto-optimal extraction per iteration is used as a baseline for work efficiency. It is possible to extract the top $n$ labels keeping the priority order intact but with no guarantee that labels after the first extraction are globally Pareto-optimal. Figure 4 shows the relative decrease in work efficiency as the number of label extractions per iteration increases. As the number of objectives increases, this multi-pop strategy extracts more Pareto-Optimal labels, thereby reducing the impact on work inefficiency.

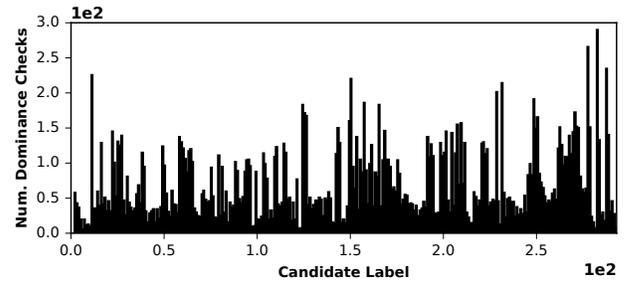

(a) Two Objectives

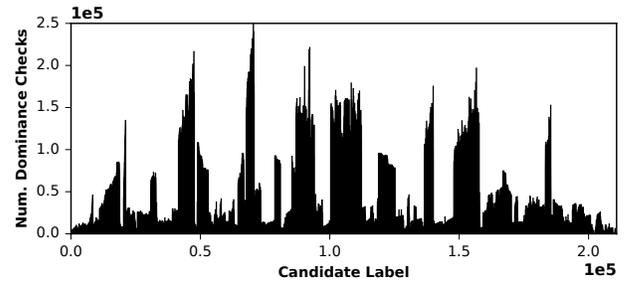

(b) Six Objectives

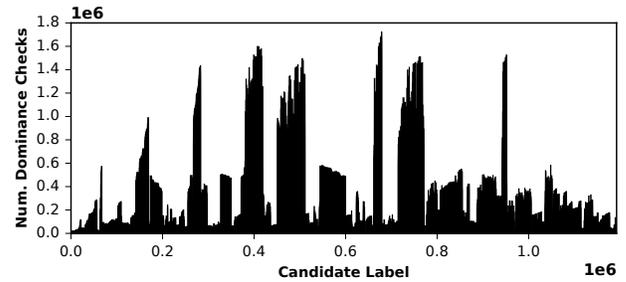

(c) Twelve Objectives

Figure 3: Distribution of per-label comparisons performed in sequential MOS for Route 1 for two, six, and twelve objectives. From two to six objectives, the number of dominance checks and labels processed go up by three orders of magnitude, and from six to twelve, they increase by an additional order of magnitude.

MOS is unique in that it performs significant and variable work for each label, making it sensitive to work efficiency. In general, graph algorithms have only a few (tens) instructions per node to process, where extra iterations are acceptable when they come with the potential for higher orders of parallelism. In MOS, the work for each label is significant, therefore, wasted label computations have a considerable impact on performance (see Figure 2, where execution time scales three orders of magnitude higher than the number of labels extracted). A first-in-first-out (FIFO) queue is evaluated to explore the potential impact of relaxing priority for label extractions. Figure 4 (right) shows the work efficiency of a FIFO queue normalized to the sequential baseline that uses a priority queue (PQ) for OPEN. Note that at two objectives, over 6× the number of labels are extracted with FIFO compared to PQ, exceeding the limits of the chart. In all cases, even at the maximum objectives, over 25% additional work is performed. Since labels are computationally expensive at increasing numbers of objectives, any increase in work needs to be compensated with significant reductions in queue extractions and updates. If the queue operations are not the bottleneck (as suggested by Figure 2), then PQ is preferred. The general trend is that relaxing the priority queue results in extractions farther from the global Pareto front, indicating the importance of maintaining order in label processing.

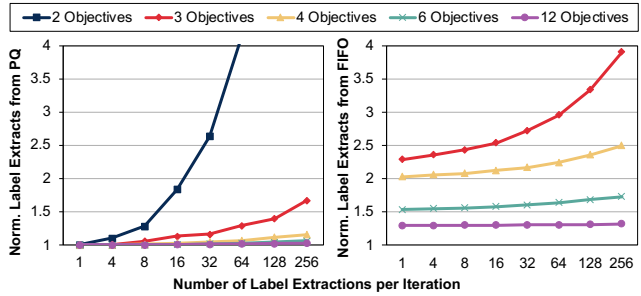

**Figure 4: TMPLAR Route 1 sweep of label extractions from a priority queue (PQ) and FIFO normalized to a PQ with single extraction per iteration.**

## 5 Ordered Parallel MOS

In MOS, labels are expensive, and redundant computations impact work efficiency. Deviation from the priority order results in a significant increase in the number of labels. Ordered Parallel Multi-Objective Shortest-Path (OPMOS) builds upon MOS in Alg. 1, introducing parallel computations in a load-balanced manner while maintaining as close to the global priority of label extractions as possible. OPMOS proposes a parallel multi-threaded execution model that operates on a global priority queue. At a high level, the main thread extracts several labels (determined by the NUM_POP system parameter) from the queue of prioritized labels (OPEN) in lexicographic order. These labels are distributed among worker threads for parallel execution using a MOS-centric

---

**Algorithm 2** OPMOS with NUM_POP & NUM_THDS parameters

**Input:** Edge costs $C$, heuristics $H$, start $v_s$ and goal $v_g$ nodes
1: ShMem Data Structures: OPEN, $G_{OP}$, $G_{CL}$, $P$, $cB_{reg}$, $cB_{goal}$, $nB_{reg}$, $nB_{goal}$, $P_{INS}$, $P_{DEL}$, OPEN$_{INS}$, OPEN$_{DEL}$, $G_{CL_{DEL}}$, $UpdatesRdy$
2: $tid \leftarrow$ Thread ID; #WT $\leftarrow$ NUM_THDS$-1$
3: **if** Main **then** ▷ Main Thread
4:     $l_s \leftarrow (v_s, \hat{0})$
5:     OPEN.insert($l_s$), $G_{OP}(v_s)$.insert($l_s$)
6:     **while** ($cB_{reg}$.len + $cB_{goal}$.len + $nB_{reg}$.len + $nB_{goal}$.len) > 0 **do**
7:         Barrier Synchronization
8:         $nB_{reg} \leftarrow \emptyset$; $nB_{goal} \leftarrow \emptyset$
9:         **while** ($nB_{reg}$.len + $nB_{goal}$.len) < NUM_POP **and** OPEN $\neq \emptyset$ **do**
10:           $l$ = OPEN.popmin()
11:           **if** PrunedFrom($l$, $G_{OP}(v(l))$) **then continue**
12:           $G_{OP}(v(l))$.remove($l$); $G_{CL}(v(l))$.insert($l$)
13:           **if** $v(l) = v_g$ **then**
14:               $nB_{goal}$.insert($l$)
15:           **else**
16:               $nB_{reg}$.insert($l$)
17:         NumReturned $\leftarrow$ 0
18:         **while** NumReturned < #WT **do**
19:           **for** $tid$ from $1\ldots$#WT **do**
20:               **if** $UpdatesRdy[tid] = 1$ **then**
21:                   ApplyUpdates($tid$, $P_{INS}$, $P_{DEL}$, $G_{CL_{DEL}}$)
22:                   ApplyUpdates($tid$, OPEN$_{INS}$, OPEN$_{DEL}$)
23:                   $UpdatesRdy[tid] \leftarrow 0$
24:                   NumReturned $\leftarrow$ NumReturned $+1$
25:     Barrier Synchronization
26:     **return** $P$
27: **else** ▷ Worker Thread (WT)
28:     **while** true **do**
29:         Barrier Synchronization
30:         **if** ($cB_{reg}$.len + $cB_{goal}$.len + $nB_{reg}$.len + $nB_{goal}$.len) > 0 **then**
31:           **return**
32:         **for** $bid$ from $1\ldots size(cB_{goal})$ **do**
33:           $l \leftarrow cB_{goal}[bid]$
34:           **for** $i = (tid + bid) \% (\#WT); i < \#Nodes; i+= \#WT$ **do**
35:               Prune$G_{OP}(i, l, $OPEN$_{DEL})$
36:           **if** ($bid \% \#WT) = tid$ **then**
37:               PruneAndInsertP($l$, $P_{INS}$, $P_{DEL}$)
38:         $s_l, s_{nbr}, e_l, e_{nbr} \leftarrow$ NbrSplitting($cB_{reg}$, $tid$)
39:         **for** $l'$ in range($s_l, s_{nbr}, e_l, e_{nbr}$) **do**
40:           ProcessRegularLabel($l'$, OPEN$_{INS}$, OPEN$_{DEL}$, $G_{CL_{DEL}}$)
41:         $UpdatesRdy[tid] \leftarrow 1$

---

load-balancing scheduler. The parallel execution of labels is done asynchronously with label extractions from OPEN. The worker threads perform the dominance and pruning checks and create updates for the processed labels. These updates are then applied asynchronously in the main thread to hide the sequential latency of these operations.

Algorithm 2 presents the pseudo-code for OPMOS. The data structures remain unchanged from Alg. 1. However, OPEN, $G_{OP}$, $G_{CL}$, and $P$ are initialized in shared memory for efficient parallel access by worker threads. Per-worker-thread data structures, $P_{INS}$, $P_{DEL}$, OPEN$_{INS}$, OPEN$_{DEL}$, and $G_{CL_{DEL}}$ are also initialized in shared memory to track local updates during parallel execution of labels. The updates are communicated between the worker and main threads using the $UpdatesRdy$ shared memory flags. The label extractions are performed in the main thread from OPEN and stored in the *bag* data structures, described in more detail next.

## 5.1 Asynchronous Execution Model

OPMOS leverages a single centralized priority queue to maintain as close to a global priority order as possible and reduce work inefficiency. The MOS characterization suggests that many labels can be extracted without significantly impacting work efficiency. However, applying the label updates in the main thread creates a sequential bottleneck. To overcome this challenge, OPMOS proposes pipelining label extractions and updates across consecutive iterations. A set of labels extracted in a given iteration is processed in the subsequent iteration while a new set of labels is concurrently extracted from the priority queue. Moreover, as labels are processed in worker threads, their updates are applied asynchronously in the main thread. This enables OPMOS to hide priority queue operations by performing label processing asynchronously. As the complexity of label processing increases with the number of objectives, the asynchronous model has more opportunities to hide sequential bottlenecks.

To enable asynchronous execution, OPMOS proposes to differentiate between the bag of labels being processed on the current iteration and the bag of labels being extracted for the next iteration. The bag of labels processed by the worker threads in a given iteration are stored in the $cB_{reg}$ and $cB_{goal}$ data structures, and the labels that are extracted for the next iteration are stored in the $nB_{reg}$ and $nB_{goal}$ data structures. From an implementation perspective, the physical bag data structures are logically swapped on each iteration and are independently managed by the main and worker threads. After initialization, OPMOS diverges into a main thread and multiple (NUM_THDS −1) worker threads. The main thread creates the start label (line 4) and inserts the label into both OPEN and $G_{OP}$ (line 5) to initialize the system.

At any given iteration *i*, the main thread performs label distribution and updates. While the main thread executes these operations, the worker threads process labels from the bags extracted in the previous iteration $i − 1$, i.e., $cB_{goal}$ and $cB_{reg}$ (lines 32-40). This allows decoupled execution of label computations from the priority queue operations. At the start of iteration *i*, the main thread sequentially extracts labels from OPEN and $G_{OP}$ and inserts the labels into $G_{CL}$ (lines 10, 12). If a label corresponding to the goal node, $v_g$, is extracted (line 13), then the label is inserted into the next iteration goal-node bag $nB_{goal}$ (line 14). Otherwise, the label is inserted into the next iteration bag for regular labels, $nB_{reg}$ (line 16). Label extractions continue until either NUM_POP labels are extracted or OPEN is empty (line 9).

After each worker thread finishes performing its necessary work and preparing its updates, it signals to the main thread that the updates are ready (line 41). The main thread polls the worker threads to check if their updates are ready (lines 18-20). It sequentially processes the updates from worker threads asynchronously (lines 21-22) and resets the worker status flag (line 23). This allows update latency to be hidden if the runtime of worker threads exhibits variability in label computations (cf. Section 4). The main thread does not progress to the next iteration until all the worker threads return and their updates are applied. When performing label updates on data structures that are visible to the worker threads for label processing ($G_{OP}$, $G_{CL}$, and $P$), care must be taken to ensure that all label metadata is consistently propagated among threads. This is ensured by tracking each label's metadata with a ready flag. Setting this flag indicates that metadata is visible in shared memory, while clearing this flag indicates a label's deletion. The label update time is primarily spent on priority queue operations (OPEN inserts and deletes, line 22). Note that OPEN is a local data structure visible only to the main thread.

OPEN inserts must be performed when the updates are applied. However, deletes can be postponed until the labels are extracted from OPEN at a later iteration. Rather than deleting from the priority queue at the time of update (in-place), deletes from OPEN are marked for removal in the relevant $G_{OP}$ entry. On OPEN extraction, the label is checked against $G_{OP}$ and discarded if it is marked for removal (line 11). This replaces the expensive random-access deletes from OPEN with cheaper priority queue extractions, reducing the sequential latency of OPEN deletes. There is no delay in a delete propagating its effect to other labels because the $G_{OP}$ entry is marked at update time, not OPEN extraction.

After all threads finish execution, they synchronize on a barrier to ensure data consistency across iterations (lines 7 and 29). Execution continues until the current and next bags are empty (line 6). Upon the main thread exiting the loop, it performs barrier synchronization to allow the worker threads to continue and observe the termination condition (lines 29-31). The main thread returns *P*, the set of exact Pareto-optimal solutions.

A key challenge for OPMOS is the latency of priority queue operations. OPMOS proposes asynchronous OPEN extractions and updates, and on-the-fly OPEN deletes to reduce the impact of the sequential latency. An alternative is to perform in-place OPEN deletes by directly removing labels from the priority queue. Similarly, extractions and updates can be performed synchronously with label processing. Processing multiple labels on a single iteration may introduce candidate labels that dominate each other or create duplicate updates.

OPMOS: Ordered Parallel Algorithm for Multi-Objective Shortest-Paths

The worker threads can synchronize to perform dominance and duplicate checks. Although this may reduce the update volume, it serializes all updates to the main thread, which results in inherently synchronous updates. To evaluate the efficacy of the proposed asynchronous execution model, these alternatives are evaluated in Section 7.

## 5.2 Label-Aware Load Balanced Execution

On each iteration $i$, the worker threads process labels in the bags $cB_{reg}$ and $cB_{goal}$ generated in the previous $i-1$ iteration. A naïve load balancer distributes the labels in the bags among workers for parallel execution. However, labels at the goal node, $v_g$ may have more complexity than regular labels. The full-index search in PruneOPEN (lines 9-10 in Alg. 1) is expensive. If these searches are not handled in parallel and performed alongside other labels, it leads to a significant load imbalance. On the other hand, if predominantly goal node labels are processed in iterations and the amount of exploitable parallelism is limited in them, then it is beneficial to distribute these labels alongside other labels. OPMOS treats goal node labels separately from other labels for balanced work execution. This splits the load-balancing problem into two parts: regular labels and goal-node labels.

Within goal-node execution (lines 9-13 in Alg. 1), there are two primary components: Pruning from OPEN and $G_{OP}$ (lines 9-10 in Alg. 1), and computations on $P$ (lines 11-13 in Alg. 1). OPMOS distributes these components separately. The PruneOPEN function in Alg. 1 performs a full-index search over OPEN. However, OPEN and $G_{OP}$ store the same labels, allowing either to be searched to determine the labels for pruning. $G_{OP}$ is organized on a labels-per-node granularity, which allows the full-index search to be spread across multiple worker threads using *node-centric* distribution. For each label in the $cB_{goal}$ bag, the nodes are split using a round-robin approach, and $G_{OP}$ for each respective node $i$ is pruned against the goal label with the Prune$G_{OP}$ function (lines 34-35). Note that a naïve round-robin approach where the nodes are split in a static way for each label can suffer from load balancing challenges when there is insufficient parallelism due to few labels in OPEN/$G_{OP}$ per node, or high load imbalance due to a few nodes with a high number of labels. To overcome this challenge, the start position for the round-robin is biased by the bag index $bid$ of the label. This allows work to be evenly distributed across the worker threads, reducing load imbalance.

For computations on $P$, the prune and insert operations require that labels are inserted after pruning. Otherwise, a label may be incidentally pruned by its pruning check in another thread. Hence, exploiting parallelism within the computations on $P$ requires fine-grain thread synchronization. OPMOS avoids this additional synchronization burden and performs $P$ computations by distributing each label in $cB_{goal}$ to an independent thread. The goal labels are assigned round-robin among the worker threads (lines 36-37).

The regular labels are processed after the goal bag computations are complete. A naïve load balancer distributes each label to a worker thread. However, the amount of work performed for each label has high variability due to an irregular number of neighbors for each node and the number of other candidate labels to check for dominance (cf. Section 3). This variability results in significant load imbalance. Therefore, OPMOS proposes a label-centric approach that distributes chunks of neighbors (i.e., candidate labels) among worker threads. The NbrSplitting function (line 38) performs this distribution for the $cB_{reg}$ bag labels. The cost-per-thread is computed as the ratio of the total number of expanded labels over the number of worker threads. The NbrSplitting function uses this cost and its thread ID to determine the chunk of labels to process. The function performs a search to determine the start ($s_l$) and end ($e_l$) labels as pointers to the $cB_{reg}$ bag, as well as the neighbor to start processing for the start label $s_{nbr}$ and to end processing for the end label $e_{nbr}$. This schedule of labels is then processed (lines 39-40), and the updates are accumulated in the relevant update data structures. Note that the ProcessRegularLabel function refers to lines 16-31 in Alg 1.

A key challenge for OPMOS scaling is the computational load variability for the parallel processing of labels. Several alternative load-balancing strategies are considered to evaluate the efficacy of the proposed label-centric approach. A naïve approach does not differentiate goal and regular labels and distributes them equally among the worker threads. Going further, the variability in regular label processing may be reduced by distributing regular labels at the neighbor level, again considering goal labels the same as regular labels. When considering goal and regular labels independently, an alternative approach is to distribute each goal label one at a time with a worker thread performing computations on $P$ and the remaining threads executing the OPEN and $G_{OP}$ pruning operations at the node granularity. These alternative approaches are evaluated against OPMOS in Section 7. Future work can supplement the proposed label-centric approach with runtime tracking of label complexity to reduce load imbalances. One can go further, breaking the dominance checks at a finer granularity. However, unlocking such parallelism may require complex serialization among threads.

## 6 Methods

The NVIDIA GH200 [19] Superchip's CPU is used for evaluation. It integrates 72 Neoverse V2 Arm cores operating at $3.1 GHz$ in a single chip. Each out-of-order core has six integer and four floating-point execution units. The memory hierarchy supports $64KB$ L1 instruction and data caches, $1MB$ private L2 cache per core, a shared $114MB$ last-level cache, and $\sim 500GB$ of on-package LPDDR5X unified memory with $512GB/sec$ memory bandwidth. Although the Superchip includes a Hopper NVIDIA H100 GPU interconnected with the CPU using NVLink, it is not used in this paper.

| # | Obj. | # | Objective | # | Objective |
|---|------|---|-----------|---|-----------|
| 1 | Distance | 5 | Vert. Acceleration | 9 | Wave Height |
| 2 | Fuel | 6 | Horiz. Acceleration | 10 | Wave Period |
| 3 | Roll | 7 | Vert. Bending Moment | 11 | Rel. Wave Bear. |
| 4 | Pitch | 8 | Vert. Shear Force | 12 | Random |

**Table 1: TMPLAR Objective List. For a given $n$ objectives, the first $n$ are used for lexicographical ordering.**

| Rt. # | Origin (Long., Lat.) | Des (Long., Lat.) | Nodes | Edges | Max Obj. |
|---|---|---|---|---|---|
| 1 | Roanoke Isl., NC 75.0°W, 36.5°N | Bahamas 76.0°W, 25.0°N | 471 | 4394 | 12 |
| 2 | Alaska 144.4°W, 58.5°N | San Diego 117.6°W, 32.7°N | 1610 | 10019 | 4 |
| 3 | Alaska 144.4°W, 58.5°N | Seattle 125.6°W, 48.4°N | 461 | 2610 | 12 |
| 4 | Guam 144.8°E, 13.4°N | Sasebo 134.1°E, 31.5°N | 201 | 2476 | 12 |
| 5 | Str. of Gibraltar 7.5°W, 36.0°N | Roanoke Isl., NC 75.0°W, 36.5°N | 778 | 7787 | 6 |

**Table 2: TMPLAR State-Space Reduced Routes**

### 6.1 TMPLAR Graphs

TMPLAR is a Python-based tool that generates ship routing graphs with up to twelve objectives (as shown in Table 1). The distance objective measures the distance along a line with constant bearing to true North (the so-called rhumb line distance). Three weather and oceanic parameters are directly used as objectives: wave height, wave direction (relative to ship bearing), and wave period. These parameters are obtained from the ERA5 dataset [10], analyzed at 3-hour intervals starting January 1, 2016. The oceanographic parameters are also used to generate seven objectives: fuel consumption (based on required propulsion in calm water [11] and due to wave resistance [6]), and six ship dynamic response objectives (calculated using a nonlinear wave-load analysis [29]): roll, pitch, vertical acceleration, horizontal acceleration, vertical bending moment, and vertical shear force. A pseudo-randomly generated objective is calculated using a seed of the latitude, longitude, and time window information at each graph edge. TMPLAR allows any number of objectives to be run for a given route, where the objectives are selected in the order specified in Table 1.

TMPLAR's state-space-reduced graph routes used for evaluation are shown in Table 2. These are generated using start and end locations as inputs, along with the start date of Jan. $1^{st}$, 2016 for weather data information, minimum and maximum ship speeds of 5 and 30 knots, and a trip length of 14 days (±1 day). Furthermore, the graphs are expanded with 10 time windows per node to capture temporal weather variations. The edge weights are populated with data for the user-specified number of objectives. The graphs in Table 2 are generated using these inputs for evaluation. An 8-hour time limit is imposed on sequential NAMOA*. Routes 1, 3, and 4 complete the maximum number of objectives (12). However, routes 2 and 4 only complete 4 and 6 objectives in this time limit.

### 6.2 NAMOA* and OPMOS Implementation

The TMPLAR tool is implemented in Python 3.10.1 [35] and used to generate the inputs for NAMOA* and OPMOS (adjacency list, edge costs, and heuristic data). These inputs are then supplied to a C++ implementation of NAMOA* and OPMOS using the python `ctypes` library. The C++ code is compiled using GCC version 11.4.0.

Two types of data structures are described in Alg. 1: label sets and the priority queue. The label sets $G_{OP}$, $G_{CL}$, and $P$ are implemented as arrays of structures with user-managed dynamic array sizing. The OPEN queue is implemented as a `std::set` with a custom lexicographical ordering function. These same data structures and their implementations are used for the parallel OPMOS algorithm, with the label set arrays stored in global shared heap memory. OPMOS uses intermediate buffers for communicating work distribution with additional shared memory data structures ($cB_{reg}$, $cB_{goal}$, $nB_{reg}$, $nB_{goal}$, $P_{INS}$, $P_{DEL}$, $OPEN_{INS}$, $OPEN_{DEL}$, $G_{CL_{DEL}}$, and $UpdatesRdy$ in Alg. 2). The threads in OPMOS are spawned using the POSIX thread (`pthread`) library for efficient multi-threaded implementation with fast inter-thread communication using shared memory.

The evaluation metrics are collected using instrumented performance counters and the C++ `chrono` library with a steady clock to ensure accurate time reporting. The total number of label extractions, candidate labels explored, and final Pareto-optimal solutions are collected using performance counters. The end-to-end execution time of OPMOS measures the total runtime of the main thread while-loop in Alg. 2 (i.e., the runtime does not include the initialization of the data structures and threads). To further describe where time is spent in execution, the runtime of sequential MOS is broken into the following components: time spent in **OPEN extractions**, time spent in **updates**, and time spent in **label processing**. In parallel OPMOS, OPEN extractions and updates can be hidden by the time spent in label processing. If time can be hidden, it is first taken from OPEN extractions and then updates. For label processing, the worst-case worker thread on each iteration is considered. This runtime is further split into the time spent in **parallel label processing**, and the remaining time is reported as **communication overheads** (time spent waiting on non-blocking barriers in Alg. 2). The average time spent in each component across all worker threads is reported. The remaining time in the main thread not spent on OPEN extractions and updates or being hidden by the worker threads is added to the **communication overheads** component. The sum of all breakdowns (unhidden OPEN extractions, unhidden updates, parallel label processing, and communication overheads) equals the total runtime of the algorithm.

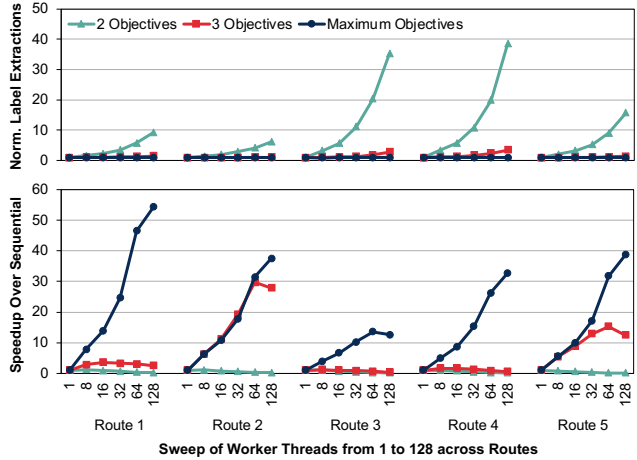

Figure 5: OPMOS speedup and labels extracted at increasing worker threads (and labels extracted) relative to sequential MOS.

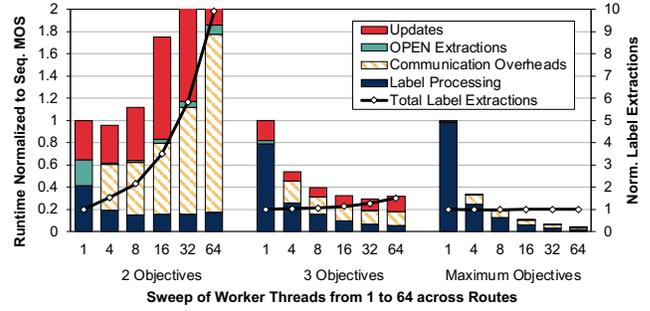

Figure 6: OPMOS geometric mean runtime with average distributions across routes for two, three, and maximum objectives.

## 7 Evaluation

OPMOS operates with two key system parameters: the number of threads (NUM_THDS) and the number of label extractions from OPEN on each iteration (NUM_POP). Figure 5 evaluates the performance of OPMOS for two (lowest), three, and the maximum number of objectives (as shown in Table 2) for each route. The number of label extractions for each iteration (NUM_POP parameter) is set equal to the number of worker threads (NUM_THDS - 1), and the number of worker threads is increased from 1 to 128. The normalized number of label extractions is plotted as a metric of work efficiency. The speedup and normalized label extractions are shown relative to sequential MOS.

The performance does not scale at two objectives due to a rapid increase in work inefficiency at higher worker threads and label extractions. The labels diverge from the global Pareto-front in the priority queue (cf. Section 4) and more redundant and unnecessary labels are processed. To gain insights into the tradeoff involved, Figure 6 shows the geometric mean runtime distribution of all routes for two objectives. The sequential baseline is dominated by priority queue operations that cause serialization bottlenecks in OPMOS. The label processing time is cut in half at four worker threads, but 50% additional labels are processed. These redundant labels make label processing time, extractions, and updates more expensive leading to runtime overheads. In addition, the label checks are relatively inexpensive (as characterized for two-objective MOS in Figure 3a). This leads to fast label processing in worker threads, which causes serialization in the main thread. The asynchronous model hides label extractions, but a significant portion of the updates and communication stalls are left unhidden. As a result, there is little performance gain (22%) at 4 worker threads. The performance scaling worsens at higher worker threads, primarily due to the rapid increase in work inefficiency.

The label processing time is observed to scale up to 64 worker threads at three objectives. This is attributed to high-quality label extractions from OPEN, leading to less work inefficiency. Moreover, there are more candidate labels per node for dominance checks at three objectives, which leads to higher complexity and processing time per label. With more work being performed by the worker threads, there are opportunities for the asynchronous model to hide the priority queue operations and the communication stalls. At 64 worker threads, the reduction in computation time leaves less room for latency hiding, and thus, the update latency begins to increase and reflect the growth in work inefficiency. This prevents OPMOS from scaling past 32 worker threads at three objectives, leading to 4× performance scaling compared to sequential MOS. The increase in label complexity and decrease in work inefficiency trends continue through maximum objectives (4 in route 2, 6 in route 5, and 12 in the remaining routes), allowing all runtime components to achieve higher scaling. This leads to a geometric mean 28× performance scaling at 64 worker threads.

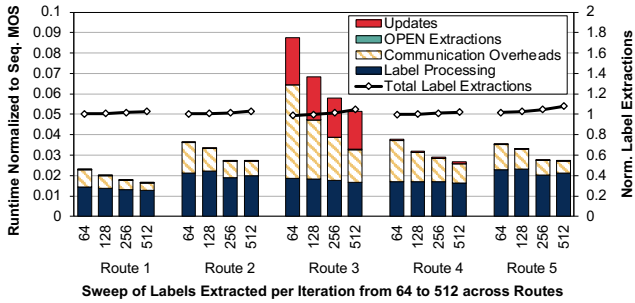

Figure 7: OPMOS runtime distributions using 64 worker threads and maximum labels extracted per iteration (NUM_POP) swept from 64 to 512.

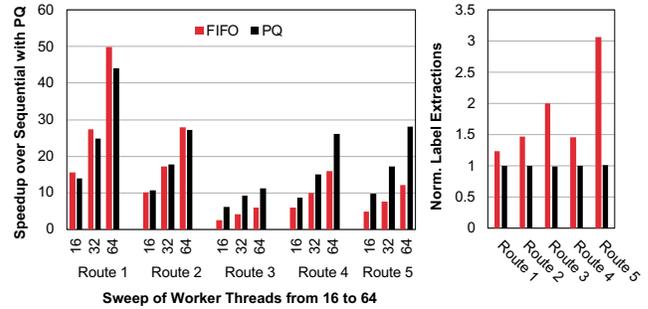

Figure 8: Comparison of OPMOS using PQ and FIFO relative to sequential MOS with PQ. Speedup is shown with increasing worker threads and label extractions (left). Normalized label extractions are shown at 64 worker threads and NUM_POP set to 64 (right).

As work efficiency is marginally impacted by the number of labels extracted at maximum objectives, it is possible to extract higher numbers of labels and further improve performance scaling. To evaluate, the NUM_POP parameter is swept from 64 to 512 labels, keeping the number of worker threads fixed to 64. Figure 7 shows the runtime distribution of OPMOS for this study, with runtime distributions normalized to sequential MOS. Note that the figure is capped at $1/10^{th}$ the sequential runtime to highlight the distributions. For all routes, the work inefficiency trends remain unchanged until a higher number of label extractions. Consequentially, the runtime components' scale and performance improves up to 256 label extractions, saturating at 512. Route 3 is the clear exception, with worse scaling and runtime dominated by updates and communication. In this route, the priority queue grows slowly, leading to relatively few labels per node (data not shown). As the number of label extractions increases, so does the time to extract from OPEN (and the time spent in updates). With 256 labels extracted, OPEN extractions become visible (hidden in all other cases). This leaves no time to hide updates and communication stalls among the threads that return in close temporal proximity. Despite these challenges, the speedup from 64 to 512 label extractions increases from 13.5× to 20×. Route 4 is similar in size to Route 3 (Table 2) but has higher density, leading to more active labels in OPEN, thereby increasing the complexity of label computations. The higher complexity labels allow OPMOS to hide OPEN extractions and updates and reduce the impact of communication stalls. As a result, most of the OPEN extractions and updates are hidden, leading to an increase in performance scaling of 26× to 37× from 64 to 512 label extractions. Routes 1, 2, and 5 are large and achieve good performance scaling. However, routes 2 and 5 operate at a much lower number of objectives (4 and 6 respectively) compared to route 1 which operates at 12 objectives. The higher complexity of label processing in route 1 leads to a 61× speedup (near-linear scaling) at 512 extractions. However, routes 2 and 5 show somewhat diminished speedups ranging from 28× to 37×, which tapers at 256 label extractions. All subsequent evaluations use 64 worker threads and 256 label extractions, where a geometric mean speedup of 34× is achieved for the maximum number of objectives.

The achieved speedups are attributed to the following OPMOS contributions: close-to-priority extractions, asynchronous execution, and the load-balancing scheme. The close-to-priority extractions allow many labels to be processed in parallel while not significantly impacting work inefficiency. The asynchronous model hides the latency of OPEN operations and communication stalls. The load balancer accounts for the diversity in label complexity to harness the variability for efficient parallel execution. The efficacy of these contributions is evaluated next.

### 7.1 Evaluation of Priority Queue vs. FIFO

OPMOS relies on processing high-priority labels to achieve work-efficient parallel execution. However, priority queue operations are expensive and lead to sequential bottlenecks. An alternative is to use a faster data structure that manages candidate label ordering in first-in-first-out (FIFO) order. FIFO inserts and deletes are cheaper than a priority queue (PQ). However, it may result in low-priority labels being extracted on each iteration. Figure 8 evaluates the performance scaling trends by sweeping worker threads and label extractions from 16 to 64 at the maximum number of objectives. The completion time of OPMOS with FIFO and PQ is normalized to the sequential MOS using PQ. Both PQ and FIFO scale with an increasing number of worker threads. However, a consistently higher number of labels are extracted and processed using FIFO, which decreases work efficiency compared to the PQ for all routes. This results in performance scaling reduction for routes 3, 4, and 5 with up to a 2× decrease in performance for FIFO at 64 worker threads.

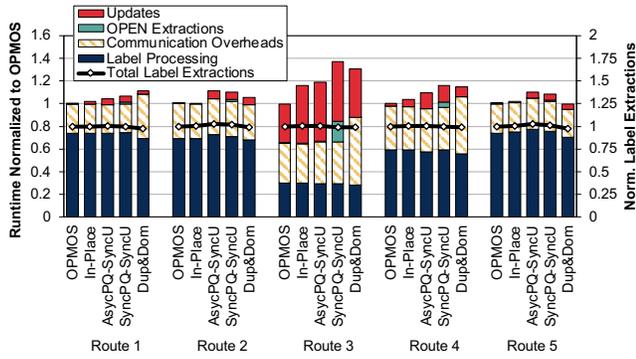

Figure 9: OPMOS runtime distribution comparison to execution model variants.

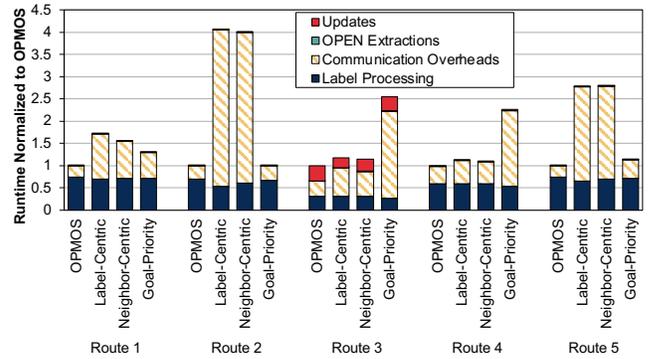

Figure 10: OPMOS runtime distribution comparison to alternative load balancing strategies.

In routes 1 and 2, although more labels are processed using FIFO, the work done per label decreases compared to the PQ. This is attributed to the order in which labels are processed. In the FIFO order, the low-priority labels are processed when inserted earlier than higher-priority labels. Such labels may perform less complex dominance checks if processed in earlier iterations. Consequently, routes 1 and 2 observe some performance gains using FIFO compared to the PQ. Overall, the priority queue yields a 47% speedup over FIFO, highlighting the importance of maintaining a close-to-priority order of label processing in OPMOS.

## 7.2 Evaluation of Asynchronous Execution

A key tenet of OPMOS efficacy is its ability to hide the sequential latency required to maintain a global priority queue. Figure 9 evaluates the runtime distributions and label extractions for alternative configurations normalized to the OPMOS asynchronous model with on-the-fly OPEN deletes. OPMOS with in-place priority queue deletes (In-Place) has the same work efficiency as OPMOS, however, it adds to the update latencies. This confirms that delaying OPEN deletes helps the asynchronous model to hide serial update operations. OPMOS with synchronous updates (AsycOPEN-SyncUpd) waits for all worker threads to return for an iteration, which prevents any update latencies from being hidden. Moreover, the updates do not propagate to active labels being processed in the given iteration, which leads to a slight decrease in work efficiency (routes 2 and 5) and an increase in label processing time. Disabling the asynchronous model (SyncOPEN-SyncUpd) introduces OPEN extraction time into the distribution. Work efficiency may improve since label extractions and processing are done in the same iteration. This leads to a slight improvement in label processing time compared to AsycOPEN-SyncUpd. However, the serialization of label extractions and updates leads to diminished performance compared to OPMOS. Finally, inter-thread duplicate and dominance checks in OPMOS (Dup&Dom) increase communication and update stalls. All worker threads need to communicate to perform their update reduction operations. This prevents updates from being sent asynchronously to the main thread, resulting in unhidden update stalls. The communication stalls also increase due to additional inter-thread synchronizations. The advantage of reducing the update volume (or unnecessary updates) is insufficient to overcome these overheads, resulting in diminished performance compared to OPMOS. Overall, the asynchronous model introduced in OPMOS efficiently hides label extractions and updates while allowing the system to minimize communication stalls.

## 7.3 Evaluation of Load Balancing Methods

The high variability in label complexity presents a challenge for efficient load distribution. Figure 10 evaluates the runtime distribution for alternative load-balancing models normalized to OPMOS. A naïve scheduler distributes all extracted labels (regular and goal) among threads (Label-Centric). The Neighbor-Centric scheduler further distributes the regular labels at their neighbor granularity. The variability in label complexity leads to significant load imbalances in both schedulers, increasing communication stalls. Routes 2 and 5 stand out due to the disproportional complexity of goal node labels compared to regular labels. The Goal-Priority scheduler handles goal labels independent of regular labels and distributes them one by one among worker threads by splitting their pruning checks at node granularity. However, the computations for $P$ are assumed to be cheap and thus executed in a single worker thread. This scheduler reduces load imbalance and nearly matches OPMOS for routes 2 and 5 but introduces significant load imbalance in routes 3 and 4. These routes do not have enough work for goal nodes to exploit the available parallelism, and their not-so-cheap $P$ computations are serialized in a single worker thread. This leads to high variability in work per thread, leading to communication stalls. OPMOS executes all goal node labels in parallel, where pruning checks and computations on $P$ are distributed among worker threads. Consequently, all routes observe low communication stalls in OPMOS.

| Route # | Max Obj. | Sequential Time (s) | OMPOS Time (s) | OPMOS Speedup |
|---|---|---|---|---|
| 1 | 12 | 2,725 | 48.22 | 57× |
| 2 | 4 | 23,083 | 628.36 | 37× |
| 3 | 12 | 0.92 | 0.054 | 17× |
| 4 | 12 | 25.26 | 0.732 | 34.5× |
| 5 | 6 | 6,968 | 190.88 | 36.5× |

Table 3: TMPLAR Pareto-Optimal execution time results for all routes. OPMOS results shown with 64 worker threads and 256 labels extracted per iteration.

### 7.4 Summary of Evaluation

MOS optimizes all objectives simultaneously and finds a set of Pareto-optimal (non-dominated) solution paths. As the number of objectives increases, so does the number of Pareto-optimal solution paths and execution times. OPMOS maintains the exact solution set while extracting maximum parallelism for acceleration. The total number of solutions obtained from the sequential MOS match perfectly with OPMOS for all experiments discussed in this paper. Furthermore, as shown in Table 3, OPMOS exploits label-level parallelism to accelerate the MOS problem by an order of magnitude or more. This allows decision-makers to make informed decisions in high-impact application scenarios.

### 8 Future Directions

A complementary approach is to reduce the complexity of MOS by adopting approximate algorithms. For example, an $\epsilon$ parameter has been proposed to influence the dominance checks [37] to enable the pruning of labels within an $\epsilon$-bounded range. Approximating the dominance checks decreases the number of labels processed and their complexity. However, the gains in performance through parallelism and work-efficiency improvements may come at the cost of solution quality. This raises concerns about the explainability of approximation techniques to accelerate OPMOS further. Therefore, on the **algorithmic front**, future research must devise methods to understand the relationship between parallelism, work efficiency, and solution quality.

To unlock parallelism further, an approach that favors parallel execution by relaxing ordered label processing can be implemented where a large number of candidate labels are processed in parallel until the algorithm settles on a set of Pareto-optimal solutions. This approach will require massively parallel hardware with fast communication to support label-level checks. Vector architectures, such as modern GPUs, and multi-node high-performance computing (HPC) clusters are potential architectures that enable such a paradigm. However, on the **architectural front**, future research must devise work-efficient and load-balanced parallel execution to unlock the performance potential of MOS.

Although this paper uses TMPLAR's ship routing application as a benchmark for MOS, more research is needed to support routes with increasing graph sizes and diverse applications. MOS can benefit real-world applications, such as autonomous systems, road networks, energy grids, and social networks, to name a few. Dealing with increasing objective counts requires state space reductions and pre-processing of graph data, akin to TMPLAR, that generates Pareto-optimal solutions within a time limit. The **benchmarking front** requires future research to create open-source MOS benchmarks for broader community adoption.

### 9 Conclusion

This paper explores the NP-hard, multi-objective shortest path problem. State-of-the-art MOS algorithms maintain a set of partial paths at each node and perform ordered processing to ensure that Pareto-optimal solution paths are generated. Finding a set of exact solutions becomes computationally intractable as the number of objectives increases. The performance characterization of MOS reveals that the computational complexity grows substantially with the number of objectives. However, due to long computational paths in label processing, there is potential for parallelism. It is concluded that ordered processing of candidate labels is needed for work-efficient parallel execution. Using this insight, the Ordered Parallel MOS (OPMOS) algorithm is proposed to handle large numbers of objectives. Novel approaches are proposed to create load-balanced and asynchronous execution of labels. The evaluation using a 72-core Arm CPU shows a geometric mean 34× speedup over sequential MOS.

### Acknowledgments

This research was supported by the U.S. Government under a grant by the Naval Research Laboratory and the National Science Foundation.